\documentclass[letterpaper,prd,nofootinbib,preprintnumbers,showpacs,showkeys,floatfix,floats,epsf]{revtex4}
\usepackage{graphicx,amssymb,epsfig}
\usepackage{citesort}
\def\figFc{
\begin{figure}[ht]
\begin{center}
\vspace{20pt}
\leavevmode
\vbox{
 \hbox{
   \epsfxsize=0.40\hsize \epsfbox{figs/HoneFc.eps}
   \hspace{30pt}
   \epsfxsize=0.40\hsize \epsfbox{figs/ZeusFc.eps}
 }
}
\vspace{20pt}
 \caption{
Comparisons of the charm production structure function, $F_{2}^{c}$, with
the data from H1 (Fig.~\protect\ref{fig:Fc}-a) \protect\cite{H1f2c01} and
ZEUS (Fig.~\ref{fig:Fc}-b). \protect\cite{ZeusF2c} The $y$-axis is
\hbox{(Experiment-Theory)/Theory}; along the x-axis, the data points are
ordered using $x$ and $Q$ as the primary and secondary sorting variables
respectively. The error bars represent the statistical and uncorrelated
systematic errors added in quadrature. }
 \label{fig:Fc}
\end{center}
\vspace{00pt}
\end{figure}
}
\def\figCcfr{
\begin{figure}[ht]
\begin{center}
\vspace{20pt}
\leavevmode
\vbox{
 \hbox{
   \epsfxsize=0.9\hsize \epsfbox{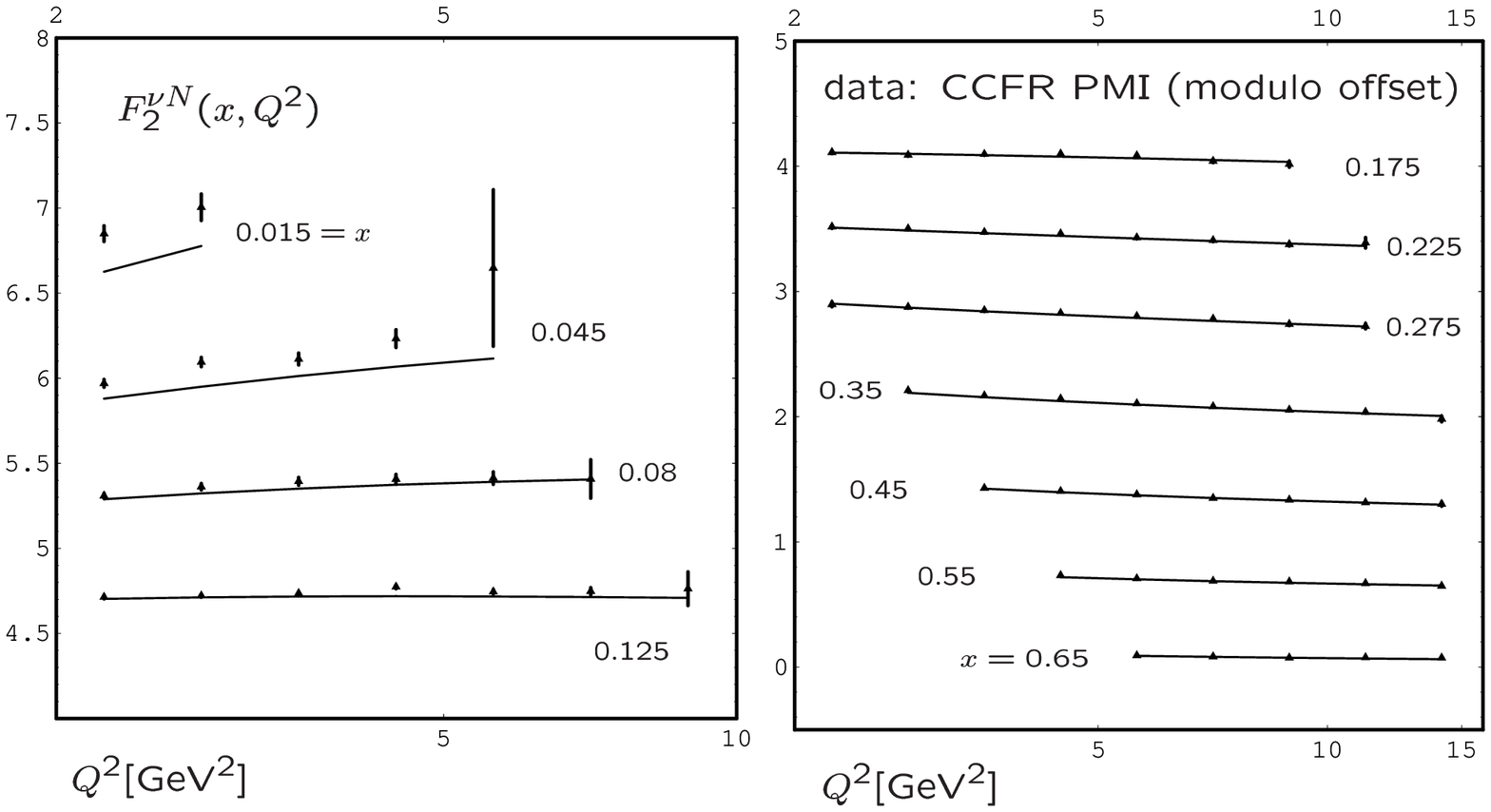}
 }
} \vspace{20pt} \caption{ CCFR $F_2^{\nu N}$ structure function (from the
{\it physics model independent} analysis, Ref.~\protect\cite{pmi}) compared
to CTEQ6HQ fit: (a) low $x$ bins (left panel) show a systematic
disagreement; (b) medium to high $x$ bins (right panel) show good agreement.} %
\label{fig:f2nu}
\end{center}
\vspace{00pt}
\end{figure}
}
\newcommand{\figUDqkA}
{
\begin{figure}[tbh]
\begin{center}
\vspace{20pt}
\leavevmode
\vbox{
 \hbox{
   \epsfxsize=0.40\hsize \epsfbox{figs/UqkQ0A.eps}
   \hspace{20pt}
   \epsfxsize=0.40\hsize \epsfbox{figs/DqkQ0A.eps}
 }
}
\vspace{20pt}
 \caption{Comparison of CTEQ5HQ, CTEQ6M, and CTEQ6HQ parton
 distributions at $Q=m_c=1.3$ GeV: (a) u-quark; and (b) d-quark. The axes are scaled
 to highlight the valence components of these distributions.}
 \label{fig:UDqkA}
\end{center}
\vspace{00pt}
\end{figure}
}
\newcommand{\figUDqkB}
{
\begin{figure}[tbh]
\begin{center}
\vspace{20pt}
\leavevmode
\vbox{
 \hbox{
   \epsfxsize=0.40\hsize \epsfbox{figs/UqkQ0B.eps}
   \hspace{20pt}
   \epsfxsize=0.40\hsize \epsfbox{figs/DqkQ0B.eps}
 }
}
\vspace{20pt}
 \caption{Same as Fig.~\ref{fig:UDqkA}, except the axes are scaled to highlight
 the sea components of the PDFs: (a) u-quark; and (b) d-quark. }
 \label{fig:UDqkB}
\end{center}
\vspace{00pt}
\end{figure}
}
\newcommand{\figGluA}
{
\begin{figure}[ht]
\begin{center}
\vspace{20pt}
\leavevmode
\vbox{
 \hbox{
   \epsfxsize=0.55\hsize \epsfbox{figs/GluQ0A.eps}
 }
}
\vspace{20pt}
 \caption{Comparison of the gluon distributions at $Q_0$= 1.3~GeV.}
 \label{fig:GluA}
\end{center}
\vspace{00pt}
\end{figure}
}
\newcommand{\figStrA}
{
\begin{figure}[tbh]
\begin{center}
\vspace{20pt}
\leavevmode
\vbox{
 \hbox{
   \epsfxsize=0.55\hsize \epsfbox{figs/StrQ0A.eps}
 }
}
\vspace{20pt}
 \caption{Comparison of the strange distributions at $Q_0$= 1.3~GeV.}
 \label{fig:StrA}
\end{center}
\vspace{00pt}
\end{figure}
}
\newcommand{\figChmGlu}
{
\begin{figure}[th]
\begin{center}
\vspace{20pt}
\leavevmode
\vbox{
 \hbox{
   \epsfxsize=0.40\hsize \epsfbox{figs/ChmQ3.eps}
   \hspace{20pt}
   \epsfxsize=0.40\hsize \epsfbox{figs/GluQ3.eps}
 }
}
\vspace{20pt}
 \caption{Comparison of the charm and gluon distribution at $Q^2 =10$
 GeV$^{\mathrm{2}}$.
}
 \label{fig:ChmGlu}
\end{center}
\vspace{00pt}
\end{figure}
}
\def\figQdep{
\begin{figure}[ht]
\begin{center}
\vspace{20pt}
\leavevmode
\vbox{
 \vbox{
   \epsfxsize=0.50\hsize \epsfbox{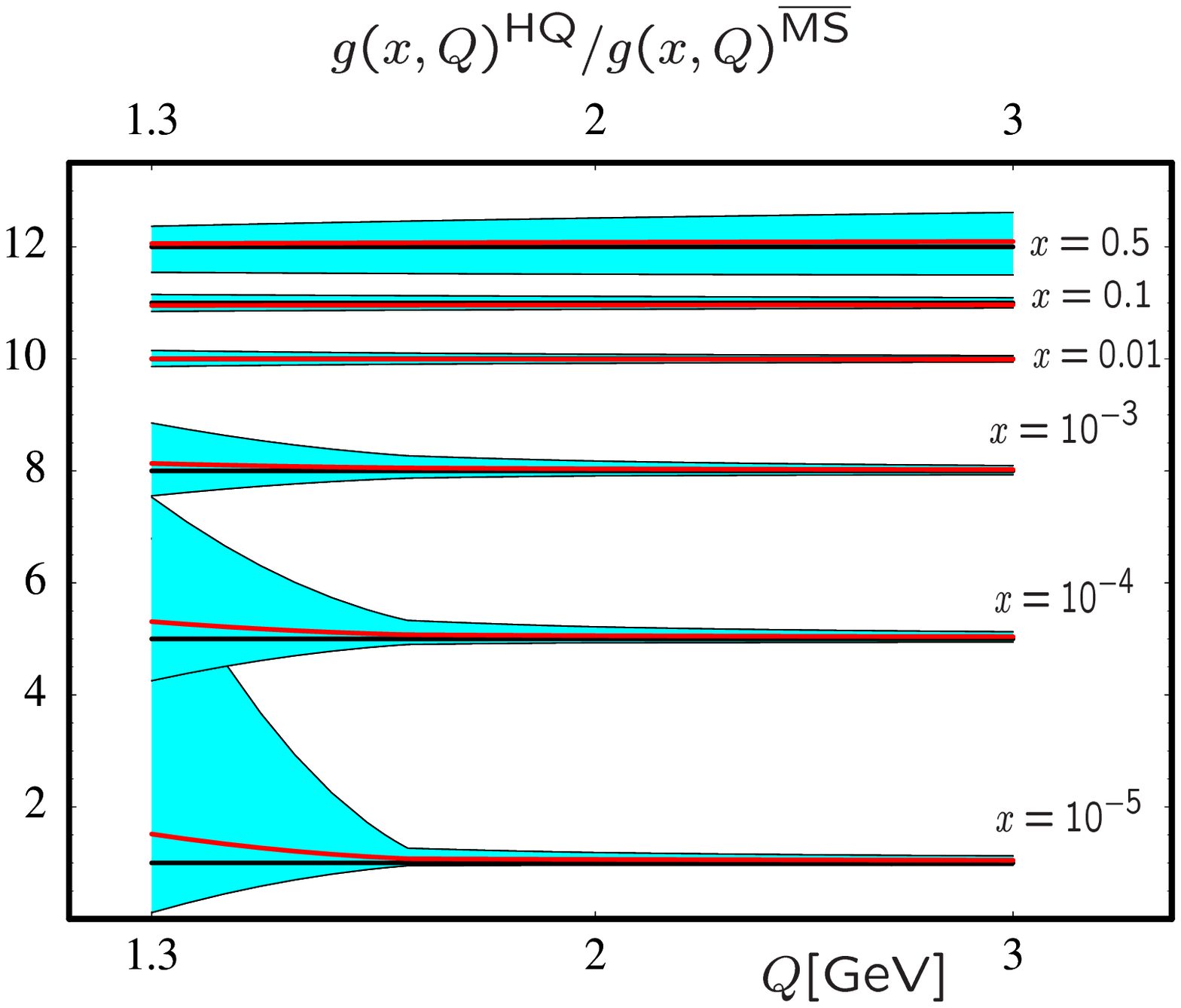}
   \vspace{20pt}
   \epsfxsize=0.50\hsize \epsfbox{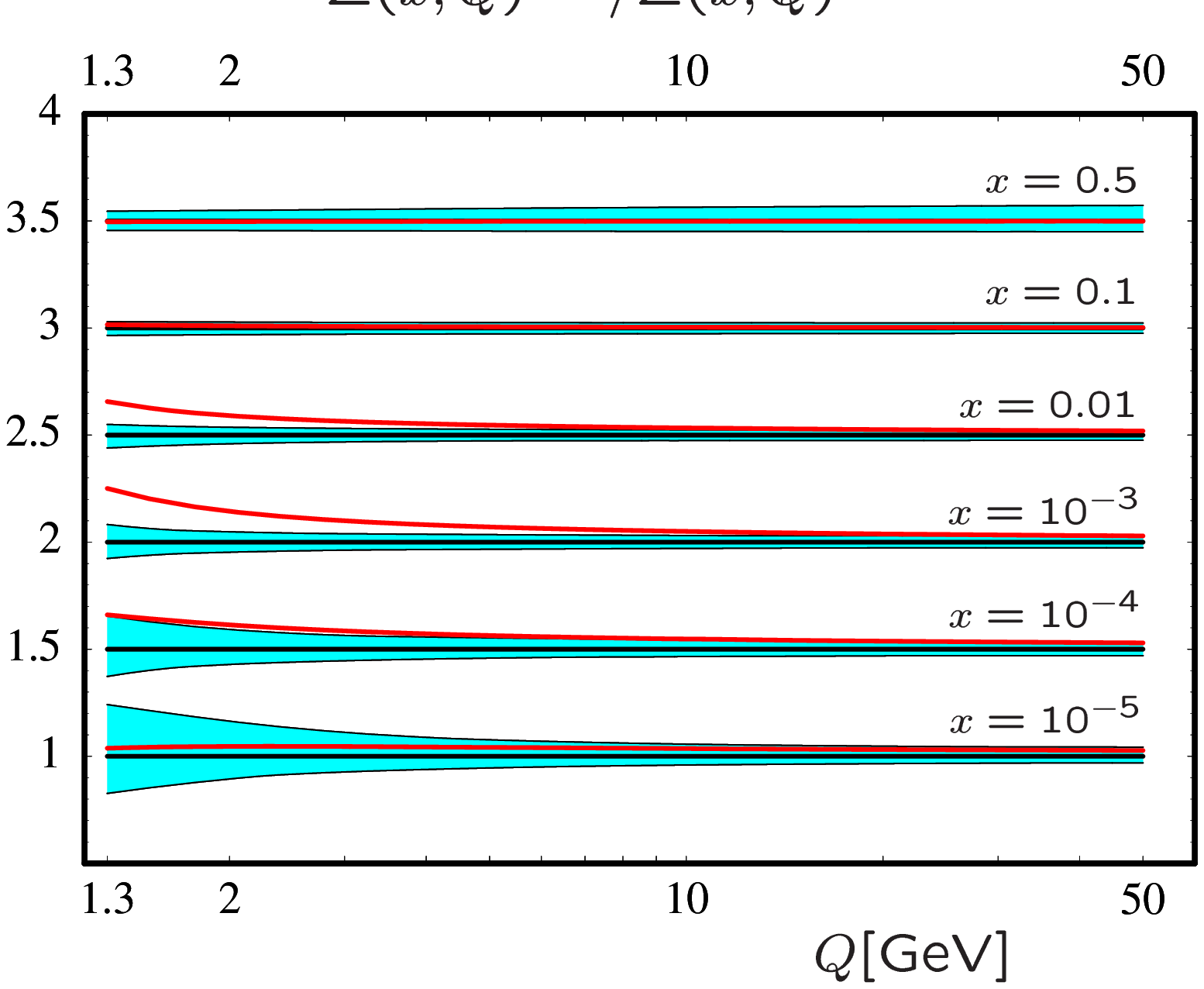}
 }
}% \vspace{20pt} 
\caption{ Ratio of GM (C6HQ) to ZM (C6M) parton
distributions as a function of $Q$ [GeV] along with the uncertainty band of
the latter. The offset on the y-axis is arbitrary; i.e.~the horizontal
lines all correspond to $\Sigma(x,Q)^{\rm HQ}
/\Sigma(x,Q)^{\scriptstyle{\overline{\rm MS}}}=1$. Plotted are ratios for
$x=\{10^{-5},\ 10^{-4},\ 10^{-3},\ 10^{-2},\ 0.1,\ 0.5\}$ where $x$ is
increasing from the lowest to the uppermost curve. We display: (a) the
gluon; and (b) the singlet quark 
$\Sigma \equiv \sum_{q} (q +{\bar q})$
\label{fig:Qdep}}
\end{center}
\vspace{00pt}
\end{figure}
}
\newcommand{\tblfits}
{
\begin{table}
\begin{center}
\begin{tabular}{|c|c|rl|rl|rl|rl|}
\hline %
Data set & \# pts & \multicolumn{2}{|c|}{CTEQ6HQ} & \multicolumn{2}{|c|}{CTEQ6M}   %
& \multicolumn{2}{|c|}{C6M$\bigotimes$GM}& \multicolumn{2}{|c|}{C6HQ$\bigotimes$ZM} \\ %
\hline
\multicolumn{1}{|l|}{Bcdms\_p} & \multicolumn{1}{|r|}{339} & 370 & (1.09) & 370 & (1.09) & 370 & (1.11) & 373 & (1.10) \\
\multicolumn{1}{|l|}{Bcdms\_d} & \multicolumn{1}{|r|}{251} & 269 & (1.07) & 279 & (1.11) & 274 & (1.07) & 281 & (1.12) \\
\multicolumn{1}{|l|}{Zeus}     & \multicolumn{1}{|r|}{104} & 94  & (0.91) & 102 & (0.98) & 258 & (2.84) & 387 & (3.72) \\
\multicolumn{1}{|l|}{H1a}      & \multicolumn{1}{|r|}{126} & 124 & (0.99) & 130 & (1.03) & 135 & (1.11) & 123 & (0.98) \\
\multicolumn{1}{|l|}{H1b}      & \multicolumn{1}{|r|}{129} & 103 & (0.80) & 111 & (0.86) & 119 & (0.84) & 104 & (0.80) \\
\multicolumn{1}{|l|}{H1c}      & \multicolumn{1}{|r|}{229} & 266 & (1.16) & 261 & (1.14) & 474 & (2.11) & 364 & (1.59) \\
\multicolumn{1}{|l|}{Nmc\_p}   & \multicolumn{1}{|r|}{201} & 304 & (1.51) & 299 & (1.49) & 273 & (1.35) & 366 & (1.82) \\
\multicolumn{1}{|l|}{Nmc\_d/p} & \multicolumn{1}{|r|}{123} & 112 & (0.91) & 111 & (0.91) & 111 & (0.90) & 114 & (0.92) \\
\multicolumn{1}{|l|}{Ccfr\_F2} & \multicolumn{1}{|r|}{69}  & 90  & (1.30) & 120 & (1.74) & 116 & (1.82) & 107 & (1.55) \\
\multicolumn{1}{|l|}{Ccfr\_F3} & \multicolumn{1}{|r|}{86}  & 35  & (0.41) & 37  & (0.43) & 36  & (0.40) & 36  & (0.42) \\
\multicolumn{1}{|l|}{E605}     & \multicolumn{1}{|r|}{119} & 102 & (0.86) & 103 & (0.86) & 101 & (0.86) & 102 & (0.86) \\
\multicolumn{1}{|l|}{Cdf\_wasy}& \multicolumn{1}{|r|}{11}  & 9   & (0.78) & 9   & (0.83) & 9   & (0.83) & 9   & (0.78) \\
\multicolumn{1}{|l|}{E866}     & \multicolumn{1}{|r|}{15}  & 5   & (0.34) & 6   & (0.43) & 6   & (0.43) & 5   & (0.34) \\
\multicolumn{1}{|l|}{D0\_jet}  & \multicolumn{1}{|r|}{90}  & 71  & (0.79) & 49  & (0.55) & 49  & (0.55) & 71  & (0.79) \\
\multicolumn{1}{|l|}{Cdf\_jet} & \multicolumn{1}{|r|}{33}  & 55  & (1.66) & 50  & (1.51) & 50  & (1.51) & 55  & (1.66) \\ %
\hline %
\multicolumn{1}{|r|}{All}& \multicolumn{1}{|r|}{1925} %
& \textbf{2008} & (1.04)&   \textbf{2037} & (1.06) & \textbf{2431} & (1.26) & \textbf{2496} & (1.30) \\ %
\hline
\end{tabular}
\end{center}
\caption{Comparison of the $\chi^2$ values of the general-mass CTEQ6HQ fit
(3rd column) with the zero-mass CTEQ6M fit (4th column). Also included
are comparisons with two ``mis-matched'' cases  when GM and ZM parton
distribution functions are convoluted with the other (wrong)
hard-cross-section (5th and 6th columns). The first number of each entry is
the $\chi^2$ value, the number  in parenthesis is $\chi^2$ per
number of points.
Correlated
systematic errors, if available, are included.}
\label{tbl:fits}
\end{table}
}

\newcommand{\plett}[3]{{Phys.\ Lett.} {\bf #1} (19#2) #3}
\newcommand{\beq}{\begin{equation}}
\newcommand{\eeq}{\end{equation}}
\newcommand{\beqa}{\begin{eqnarray}}
\newcommand{\eeqa}{\end{eqnarray}}

\def\gsim{\, \lower0.5ex\hbox{$\stackrel{>}{\sim}$}\, }
\def\lsim{\, \lower0.5ex\hbox{$\stackrel{<}{\sim}$}\, }
\newcommand{\etal} {\textit{et.\ al.}}
\newcommand{\Msbar} {{\small {$\overline {MS}$}}}

\newcommand{\DATE}
{\today}

\newcommand{\PPrtNo}
{ hep-ph/0307022\\MSU-HEP-030101\\ BNL-NT-03/2\\ RBRC-325}

\newcommand{\TITLE}
{\Large CTEQ6 Parton Distributions with Heavy Quark Mass Effects \rule{0em}{3.3ex}}

\newcommand{\AUTHORS}
{
Stefan Kretzer$^{a,b,c}$,
H.L.\ Lai$^d$,
Fredrick I.\ Olness$^e$,
W.K.\ Tung$^a$ \\
}
\newcommand{\INST}
{
$^a$Department of Physics and Astronomy, Michigan State University,\\
         East Lansing, MI 48824 USA \\
\medskip
$^b$Physics Department, Brookhaven National Laboratory,\\
Upton, New York 11973, USA \\
\medskip
$^c$RIKEN-BNL Research Center, Bldg. 510a, Brookhaven
National Laboratory, \\
Upton, New York 11973 -- 5000, USA \\
\medskip
$^d$Taipei Municipal Teacher's College, Taipei, Taiwan\\
\medskip
$^e$Department of Physics, Southern Methodist University,
Dallas, Texas, 75275 USA \\
} %

\newcommand{\ABSTRACT}%
{Previously published CTEQ6 parton distributions adopt the
conventional zero-mass parton scheme; these sets are most appropriate
for use with massless hard-scattering matrix elements commonly found in
most physics applications. For precision observables which are
sensitive to charm and bottom quark mass effects, we provide in this paper an
additional CTEQ6HQ parton distribution set determined in the more
general variable flavor number scheme which incorporates heavy flavor
mass effects.
The results are obtained by combining these parton distributions with
consistently matched  DIS structure functions computed in the same scheme.
We describe the analysis procedure, examine the predominant features of the
new distributions, and compare with previous distributions.
}

\begin{document}

\setcounter{footnote}{0}
% Standard Simple Cover Page for Papers -- wkt 1/11/95

\begin{titlepage}

\noindent
\begin{tabular}{l}
\DATE
\end{tabular}
\hfill
\begin{tabular}{l}
\PPrtNo
\end{tabular}

\vspace{1cm}

\begin{center}
                           % Title
\renewcommand{\thefootnote}{\fnsymbol{footnote}}
{
\LARGE \TITLE
%\footnote[2]{\THANKS}
}

\vspace{1.25cm}
                          % Authors
{\large  \AUTHORS}

\vspace{1.25cm}

                          % Institutions
\INST
\end{center}

\vfill

\ABSTRACT                 % Abstract

\vfill

\end{titlepage}
%%%%%%%%%%%%%%%%%%%%%%%%%%%%%%%%%%%%%%%%%%%%%%%%%%
\newpage
\tableofcontents
\newpage

%%%%%%%%%%%%%%%%%%%%%%%%%%%%%%%%%%%%%%%%%%%%%%
%%%%%%%%%%%%%%%%%%%%%%%%%%%%%%%%%%%%%%%%%%%%%%%%%%%%%%%%%%%%%%%%%%%
\section{Introduction}

Parton distributions provide the essential link between the
theoretically calculated partonic cross-sections, and the
experimentally measured physical cross-sections involving hadrons and
mesons.  This link is crucial if we are to make incisive tests of the
standard model, and search for subtle deviations which might signal
new physics.
 Since perturbative calculations are renormalization scheme
dependent,\footnote{%
We shall use the term \emph{renormalization scheme} in the general sense to
include both the usual renormalization (ultra-violet subtraction) scheme, and
the factorization (infrared and collinear subtraction) scheme.}
it is important to use properly matched hard-scattering   cross-sections
%matrix elements
and parton distribution functions (PDFs) in evaluating
factorized cross-sections for physical applications.
 This issue is particularly relevant for applications involving heavy quarks,
since the heavy quark  introduces a new mass scale
which leads to complications of the PQCD formalism.

For most physical applications, it is convenient (and a good approximation)
to use hard-scattering  cross-sections calculated in the zero parton-mass
limit. In fact, next-to-leading-order hard-scattering  cross-sections in
the \emph{non-zero quark-mass case} have only been calculated for some very
basic processes, such as deep inelastic scattering (DIS); they are not yet
available for most interesting ``new physics'' applications.  Thus, the
most useful parton distributions for general applications are those
determined in a global QCD analysis using the zero quark-mass
approximation. This was the choice made for the latest series of CTEQ6
parton distributions: CTEQ6M, CTEQ6D, and CTEQ6L.\ \cite{cteq6} For
definiteness, this scheme will be referred to as the zero-mass
variable-flavor-number scheme (ZM-VFNS).

This paper extends Ref.~\cite{cteq6} by performing a similar global
analysis using the generalized (non-zero quark-mass) \Msbar\ perturbative
QCD framework of Refs.~\cite{ColTun,AOT,ACOT,Collins98,ASTW},
which we label
the general-mass variable-flavor-number scheme (GM-VFNS). When matched to
the corresponding hard-scattering  cross-sections calculated in the same scheme, the
combination
should provide a more accurate description of the precision DIS structure
function data, as well as other processes which are sensitive to charm and
bottom mass effects. The main result we present here is a new set of parton
distributions evaluated in this GM-VFNS which we identify as
CTEQ6HQ.%
\footnote{A fit named CTEQ6F3 employing the three fixed-flavor-number
scheme (3-FFNS) for DIS (cf., Refs.~\cite{F3NLO,grs}) which is suitable for
special applications in that scheme, will be presented separately.}

Perturbative QCD with non-zero quark-masses is well-established in the
literature.\ \cite{Witten:bh,CWZ,ColTun,ACOT,Collins98,BuzaEtal,ChuvakinEtal} Its
implementation is, however, necessarily more complicated than the
corresponding zero-mass case.  In addition to the familiar
renormalization-scheme dependence, more ambiguities arise from the
implementation of the charm and bottom thresholds (in relation to their
masses), and the detailed way the 3-flavor, 4-flavor, and 5-flavor
renormalization schemes are matched at the transition scales. Several
schemes proposed in the recent literature are formally equivalent to each
other at high energies; but, they can differ in the threshold region, as
well as in the intermediate energy range, where much of the experimental
data lie.\ \cite{ACOT,BuzaEtal,ChuvakinEtal,ASTW,NasonEtal,TKS} Some of
these schemes are more natural and numerically robust, e.g. insensitive to
variations of the remaining parameters (such as scale), than others. We
follow the procedure proposed in Refs.~\cite{TKS,acotchi} which has been
shown to be particularly stable as it naturally takes into account the
dominant (logarithmic {\it and} non-logarithmic) mass-dependent kinematic
effects with a physically motivated choice of the scaling variable.

 Section~2 summarizes the method of analysis.  Because this work
represents an extension of the zero quark-mass CTEQ6 global QCD
analysis, we shall focus mostly on the new elements
related to non-zero quark mass; details that are in common with the
massless case can be found in Ref.~\cite{cteq6}.
 Section~3 describes the global analysis of data.
 Section~4 compares the new parton distributions with data.
 Section~5 makes comparisons between the CTEQ5M, CTEQ6M and CTEQ6HQ PDFs. At
asymptotically high energies, mass effects become negligible; therefore, all
generalized parton model prescriptions should approach the same zero
quark-mass limit. We show that the differences between CTEQ6M and
CTEQ6HQ PDFs indeed vanish with increasing $Q$, as expected.
 This section also contains a discussion of  applications and issues
related to uncertainties of physics predictions resulting from these parton
distributions.

%%%%%%%%%%%%%%%%%%%%%%%%%%%%%%%%%%%%%%%%%%%%%%%%%%%%%%%%%%%%%%%%%%%
\section{The Generalized \Msbar\
Scheme with non-zero mass heavy flavor partons}

The main feature of this analysis, compared to the standard zero-mass
parton CTEQ6 PDF sets, is the adoption of the generalized perturbative QCD
framework incorporating non-zero mass effects for the charm and bottom
quarks.\footnote{The top quark can, for all practical purposes, be treated
as a heavy particle, not a parton.}  Collins has shown that, to all orders
of the perturbation expansion, factorization of the DIS structure functions
holds for massive partons to the same degree of rigor as in the familiar
zero-mass case.\ \cite{Collins98} The relatively simple physical ideas
underlying the full factorization proof are described in
Refs.~\cite{ASTW,TKS,acotchi}. To implement this formalism, which we shall
refer to as the generalized \Msbar\ scheme (or GM-VFNS, as mentioned
earlier), we follow the explicit procedures outlined in the Appendix of
Ref.~\cite{ASTW}, and supplement this with the threshold prescription of
Refs.~\cite{TKS,acotchi}. The latter reference also contains a review of
alternate approaches.

The NLO DIS structure functions are given by the generic formula
(suppressing the structure function label 1,2,3),%
\begin{equation}
F(x,Q)=\sum_{a}\int \frac{dz}{z}\,f_{a}(z,\mu )\
\hat{\omega}^{a}\left(
\frac{x}{z},\frac{Q}{\mu },\frac{m_{H}}{\mu },\alpha _{s}(\mu ) \right)\;
+\; {\cal O} \left(\alpha
_{s}^{2},\frac{\Lambda ^{2}}{Q^{2}},\frac{\Lambda ^{2}}{m_{H}^{2}} \right)
\label{strfn1}
\end{equation}
where ``$a$" is the initial state parton label, $f_{a}$ is the parton
distribution function, $\hat{\omega}^{a}$ is the hard-scattering
cross-section calculated in PQCD to order $\alpha_{s}^1$, $\Lambda$ is the
QCD-lambda parameter, and $m_{H}$ (generically) represents heavy quark
masses, if present. The prescription-dependence allowed by the PQCD
formalism is associated with possible implementations of the first term on
the right-hand side of Eq.~\ref{strfn1}, within the accuracy specified by
the remainder term (which will be dropped from now on).\footnote{Note,
according to the factorization proof of Ref.~\cite{Collins98}, the
remainder term contains corrections of order $\Lambda^2/Q^2$ and
 $\Lambda^2/m_H^2$, but {\it no}  corrections of order $m_H^2/Q^2$.}
In the sum over partons in Eq.~\ref{strfn1}, contributions due to the light
quark ($u,d,s$) are standard; contributions arising from massive charm and
bottom quarks, along with that of the gluon, are treated differently, as
explained below.

%%%%%%%%%%%%%%%%%%%%%%%%%%%%%%%%%%%%%%%%%%%%%%%%%%%%%%%%%%%%%%%%%%%
\subsection{Leading contributions and the ACOT($\chi$) implementation}

To be specific, we will consider DIS neutral-current charm production
process. The leading-order process is a virtual-photon scattering off a
charm parton: $\gamma^* c \to c$. The NLO QCD corrections for this process
are the boson-gluon fusion process ($\gamma^* g \to c \bar{c}$), and the
gluon radiation process ($\gamma^* c \to c g$) as well as the
charm-initated one-loop virtual process  ($\gamma^* c \to c$).

First, let us examine the kinematic region near the charm production
threshold: $W = \sqrt{Q^2 (1/x-1)} \simeq 2\, m_c$.  To make the notation
precise, we will use the renormalization scale $\mu$ to generically label
our characteristic energy scale; separately, we use  $\mu_m$ to label the
{\it matching point}, and $\mu_t$ to label the {\it transition point}
(cf.~\cite{ASTW} and below). To implement the generalized \Msbar\ scheme,
the first step is to consider a 3-flavor scheme (appropriate for energy
scales $\mu \lsim m_c$), a 4-flavor scheme (for scales $\mu \gsim m_c$),
and choose a matching point $\mu_{m}$ (on the order of $m_{c}$) where the
two schemes are matched (i.e., where the discontinuities of
$\alpha_{s}(\mu)$ and $f_{a}(x,\mu )$ are calculated).  We choose the
conventional value for the matching point, $\mu_{m}=m_{c}$, which yields
the simplest matching conditions.\ \cite{ColTun}

Having precisely defined the 3-flavor and 4-flavor schemes, which co-exist
in the region of the charm threshold, we can still choose a
\emph{transition point}, $\mu _{t}$, where one makes the transition from
the 3-flavor to the 4-flavor scheme in the calculation of physical
quantities.%
\footnote{The distinction between the matching and the
transition points is not addressed in most papers.  This distinction is
made in Ref.~\cite{Collins98}; it is discussed at length in the Appendix of
Ref.~\cite{ASTW}.  %
As mentioned above, arguably, there is a good case for
choosing the transition point $\mu_{t}$ to be higher than the simple
matching point $\mu_{m}=m_{c}$.} The choice of the transition point is
arbitrary---its choice is  part of the definition of the composite
renormalization scheme. While there is much flexibility in the choice of
$\mu_{t}$, clearly it must lie within the overlapping region of
applicability of the 3-flavor and 4-flavor schemes---therefore, we
typically choose $\mu_{t} \sim m_c$. Although a case can be made, in
principle, for choosing $\mu_{t}$ far above $m_{c}$ (since the 4-flavor
scheme is not a natural scheme just above $m_{c}$), or for considering an
$x$-dependent transition scale, $\mu _{t}(x,m_c)$ (because the threshold
$W>2\,m_c$ depends on $x$ as well as $Q$), we make the plain choice
$\mu_{t}=m_{c}$ to simplify the calculation. This choice is reasonable {\it
only} within a prescription for handling $x$-dependent threshold effects of
Eq.~\ref{strfn1} which naturally suppresses the 4-flavor contribution to
the physical structure function when $\mu \sim m_{c}$ and/or $W \gsim
2\,m_c$. The specific renormalization prescription of Ref.~\cite{TKS} has
precisely this feature, as we shall outline below.

Detailed descriptions of the various NLO contributions to the
the generalized  \Msbar\ scheme were discussed in Ref.~\cite{ASTW}.
 While we refrain from reproducing them here, we do
need to specify the precise way the threshold effects are
implemented when heavy quarks are involved.
Consider first the numerically significant NLO
\emph{boson-gluon fusion} contribution  ($\gamma^* g \to c \bar{c}$):
\begin{eqnarray}
F^{(1)}(\gamma^* g \to c \bar{c})
&=&
\int \frac{dz}{z}\,g(z,\mu )\ \hat{\omega}_{g} \left(\frac{x}{z},\frac{Q}{\mu },%
\frac{m_{c}}{\mu } \right)  \label{nloglu} \\
&=&
\alpha _{s}(\mu )\left[ \int_{\chi }^{1}\frac{dz}{z}g(z,\mu ) \
\omega_{g}^{1} \left(\frac{x }{z},\frac{m_{c}}{Q} \right)
-\ln \left(\frac{\mu }{m_{c}} \right )
\int_{\zeta }^{1}\frac{dz}{z} \, g(z,\mu ) \,
P_{g\rightarrow c}\left(\frac{\zeta}{z}\right) \;
\omega _{c}^{0}\left(\frac{m_{c}}{Q}\right)\right]  \nonumber
\end{eqnarray}
where the first term on the right-hand side corresponds to the unsubtracted
boson-gluon-fusion diagram contribution (with the $\alpha_{s}$ factor
explicitly taken out),\footnote{We use the notation that, for each flavor,
$\omega _{a}$ denotes the parton-level cross-section {\it before} infrared
and collinear subtraction, while $\hat{\omega}_{a}$ denotes the
corresponding infra-red safe cross-section {\it after} subtraction.} and
the second term represents the subtraction term which renders
$\hat{\omega}_{g}$  infra-red safe. The variable $\chi $ is dictated by the
kinematics of the boson-gluon-fusion partonic subprocess ($\gamma^* g \to c
\bar{c}$) to be a generalized scaling (or ``rescaling") variable
\begin{equation}
\chi =x \left(1+{4 m_{c}^{2} \over Q^{2}} \right)  \   \label{chi}
\end{equation}
where $x$ is the conventional Bjorken variable. In contrast, the other
scaling variable, $\zeta $, in Eq.~\ref{nloglu} can, in principle, be
chosen arbitrarily provided the following two constraints are imposed.

\begin{enumerate}
\item $\zeta \rightarrow x$ in the high energy limit.

\item
The same $\zeta $ variable is used here as in the order
$\alpha_{s}^{0}$ (``leading-order") process   $\gamma^* c \to c$.
\end{enumerate}
In the high energy limit  ($m_{c}^{2}/Q^{2}\rightarrow 0$),
the first constraint ensures that
both  $\zeta \rightarrow x$ and  $\chi \rightarrow x$ such that
the complete boson-gluon fusion contribution is infra-red safe.
 The leading-order process $\gamma^* c \to c$ is:
\begin{equation}
F^{(0)}(\gamma^* c \to c )
=
\int \frac{dz}{z}\,c(z,\mu )\
\hat{\omega}_{c}^0 \left(\frac{Q}{\mu },\frac{m_{c}}{\mu }\right)
\
\delta\left(1-\frac{\zeta}{z}\right)
=
c(\zeta ,\mu )~\;
\hat{\omega}_{c}^0\left(\frac{Q}{\mu },\frac{m_{c}}{\mu }\right)
\ .\label{loquark}
\end{equation}
 For the second constraint, if the same $\zeta $ variable is used for
both the boson-gluon fusion process of Eq.~(\ref{nloglu}) ($\gamma^* g \to
c \bar{c}$) and the leading-order ($\gamma^* c \to c$) contribution of
Eq.~(\ref{loquark}), then the gluon subtraction term (last term in
Eq.~\ref{nloglu}) cancels the leading-order contribution in the limit $\mu
\rightarrow m_c$; this ensures that the 4-flavor formula will reduce to the
3-flavor formula, and the dominant contribution in this region will be the
unsubtracted gluon fusion term, as required by the correct physics to this
order of PQCD.

A natural choice of $\zeta$ that ensures the best behavior at \emph{both}
the high and the low energy limits is proposed by Ref.~\cite{TKS}:
\begin{equation}
\zeta =\chi \equiv x(1+4m_{c}^{2}/Q^{2})\ \ .  \label{zetachi}
\end{equation}
This choice  has the property that as the heavy quark production threshold
is approached from above ($W\rightarrow 2m_{c}$), then $\zeta \rightarrow
1$ such that the integration measure in Eq.~\ref{nloglu}
vanishes.\footnote{This is more easily seen from the equivalent formula
$\chi =1-(1-x)(1-4m_{c}^{2}/W^{2})$.} Therefore, with the choice
$\zeta=\chi$, the subtraction term in Eq.~\ref{nloglu} and the LO quark
term, Eq.~\ref{loquark}, individually vanish at threshold by kinematics. In
addition, the cancellation between these two terms due to the dynamics of
PQCD, as explained in Ref.~\cite{ACOT}, still operates.

This results in unequaled stability of the predictions in the threshold
region, compared to previous implementations of the GM-VNFS in the
threshold region \cite{ACOT,ThorneRob,ChuvakinEtal,NasonEtal}. These
usually give rise to rather prescription-sensitive behavior just above
threshold because the perturbation-expansion-inspired prescriptions
generally overlook heavy quark production kinematic requirements, while the
physically important gluon fusion term (in the threshold region) is
strongly kinematically-suppressed. The robust implementation we adopt will
be referred to as the ``ACOT($\chi $)" scheme, following
Refs.~\cite{TKS,acotchi}.%
\footnote{So named since it is a variant of the original ACOT
approach \cite{ACOT}, replacing the naive $x$ with the kinematically natural
rescaling variable $\chi $ of Eq.~\ref{chi}.} %
%

%%%%%%%%%%%%%%%%%%%%%%%%%%%%%%%%%%%%%%%%%%%%%%%%%%%%%%%%%%%%%%%%%%%
\subsection{Other contributions to the full NLO calculation}

For the complete order $\alpha_{s}$ calculation, we must also include the
real and virtual NLO quark-initiated contribution to Eq.~\ref{strfn1},
\begin{equation}
F^{(1)}(\gamma^* c \to c +X )
\ = \
\int \frac{dz}{z} \, c(z,\mu)\,
\hat{\omega}_{c}^{1}
\left(\frac{x}{z},\frac{Q}{\mu },\frac{m_{c}}{\mu }\right)
\quad .
\label{nloquark}
\end{equation}
The hard-scattering cross-section, $\hat{\omega}_{c}^{1}(x,\frac{Q}{\mu },
\frac{m_{c}}{\mu })$, has been computed for general masses \cite{nloquark}
in the  ACOT scheme \cite{ACOT}. For scales around the canonical DIS choice
of $\mu \simeq Q$, this term is numerically insignificant compared to the
terms considered above. We include it in our calculation to ensure
(perturbative) consistency.

It is also worth mentioning that the calculation of this term, which is
computation-intensive in spite of the smallness of its numerical
contribution, can be considerably streamlined by using a simplifying
feature of PQCD in the generalized \Msbar\ scheme: one can set $m_{c}=0$ in
$\hat{\omega}_{c}^{1}$ and $\omega _{c}^{0}$ in Eq.~(\ref{nloglu}) and
Eq.~(\ref{loquark}) without sacrificing accuracy. This point is discussed
in detail in Ref.~\cite{SACOT} where the numerical insensitivity of the
calculation to the choice of $m_{c}$ is explicitly
demonstrated.\footnote{The simplified scheme \cite{SACOT} can also be
recovered by investigation of the resummed perturbation series.\
\cite{acotchi}}

%%%%%%%%%%%%%%%%%%%%%%%%%%%%%%%%%%%%%%%%%%%%%%%%%%%%%%%%%%%%%%%%%%%
\subsection{Remaining scheme and scale choices}

The only remaining arbitrariness  of our prescription is now associated
with the familiar choice of the factorization (and renormalization) scale
$\mu $. The scale dependence of perturbative calculations in general arise
from imperfect matching between the terms in the truncated perturbation
series. \ Because of the near-perfect matching in the ACOT($\chi $)
prescription discussed above (in contrast to alternate choices), the
dependence of the physical structure functions on the choice of $\mu$ is
completely negligible (cf., Ref.~\cite{TKS}) throughout the energy range of
our global analysis. For definiteness, we use $\mu^{2}=Q^{2}+m_{c}^{2}$. In
summary, the quark initiated terms in the simplified ACOT($\chi$) scheme
adopted in this work are:
\begin{equation}
\label{eq:qini}
\int_{\chi}^1 \frac{dz}{z} \left.
\,c(z,\mu )\,\hat{\omega}_{c}^{0,1}\left(\frac{\chi}{z},\frac{Q}{\mu },
\frac{m_c}{\mu}= 0\right)\right|_{\mu^2=Q^2+m_c^2}
= \int_{\chi}^1 \frac{dz}{z}\, c(z,\mu )
\,C^{0,1}_{\scriptscriptstyle \overline{\rm MS}}
\left. \left(\frac{\chi}{z},\frac{Q}{\mu} \right)\right|_{\mu^2=Q^2+m_c^2}
\end{equation}
with $C^{0,1}_{\scriptscriptstyle \overline{\rm MS}}$
representing the standard coefficient
functions for massless quarks in LO and NLO, defined in the conventional
\Msbar\ scheme.

The same treatment is applied to the $b$-quark threshold region.
Here, the 4-flavor scheme is matched onto the 5-flavor scheme for
scales larger than $m_{b}$, and the production of open bottom
particles in the final state is added to the inclusive DIS structure
functions.  In principle, a similar treatment needs to be applied for
other processes included in the global analysis (currently, Drell-Yan
and jet production) using appropriate  hard-scattering cross-sections
calculated in the massive scheme.  In practice, this is not done for
two reasons: (i) NLO calculations of the hard-scattering cross-section
involving massive quarks do not yet exist for these processes; and
(ii) unlike the case of DIS, the experimental errors in these
processes are relatively large compared to the anticipated differences
between the massless and massive scheme calculations.

%%%%%%%%%%%%%%%%%%%%%%%%%%%%%%%%%%%%%%%%%%%%%%%%%%%%%%%%%%%%%%%%%%%
\section{Global Fitting and the CTEQ6HQ Parton Distributions}

The new global fitting performed in the generalized \Msbar\ formalism
follows the same procedure as that of the earlier CTEQ6 analysis
\cite{cteq6} using the zero-mass parton formalism. The data sets used
before are supplemented by: (i) the H1 $F_{2}^{e^{+}p}$ set
\cite{H1positron99} (which was inadvertently left out in
Ref.~\cite{cteq6}); and (ii) the H1 \cite{H1f2c01} and ZEUS
\cite{Zeusf2c99} data sets for the structure function $F_{2}^{c}$ with
tagged charm particles in the final state. The additional H1
$F_{2}^{e^{+}p}$ data set does not have much influence on the new analysis,
since one already obtains an excellent fit to these data just by comparing
them with predictions of CTEQ6M. The $F_{2}^{c}$ data sets are quite
relevant for this analysis since $F_{2}^{c}$ is sensitive to the charm and
gluon distributions, which are tightly coupled in the generalized \Msbar\
formalism.%
\footnote{These data sets were not used in the CTEQ6M analysis because
$F_{2}^{c}$ is not well-defined theoretically in the the zero-mass parton
formalism.} %

The parametrization of the non-perturbative parton distribution functions
at $Q_{0}=m_{c}=1.3$~GeV is the same as in Ref.~\cite{cteq6}. For this
study, we assume as usual that charm partons are entirely ``radiatively
generated" (i.e., through QCD evolution) from the starting scale $Q_{0}$
onward. This assumption is somewhat arbitrary, and it is obviously
dependent on the choice of $Q_{0}$. The possibility for having a small
component of non-perturbative charm \cite{incc} at low $Q$, and its
physical consequence will be examined separately.

As in the previous CTEQ6 analysis, correlated experimental systematic
errors are fully incorporated whenever available. The best fit obtained
with these inputs shall be called CTEQ6HQ, or C6HQ for short. A broad
measure of the quality of this fit is provided by the overall $\chi^{2}$ of
2033 for a total number of 1950 data points ($\chi^2$/DOF = 1.04). This is
to be compared to a $\chi^{2}$ of 1946 for 1811 points ($\chi^2$/DOF =
1.07) in the case of CTEQ6M (abbreviated to C6M in the following)
\cite{cteq6}. To gain a better feel of how these fits compare, we show in
Table~\ref{tbl:fits} (the 3rd and 4th columns) the overall $\chi^{2}$
values and, in parentheses, $\chi^{2}$ per data point, as well as for the
applicable individual data sets. \tblfits The total number of data points
in this head-to-head comparison (not including the charm production data
points) is 1925. The new C6HQ fit reduces the overall $\chi ^{2}$ by 29 out
of $\sim $2000 as compared to the C6M fit. The improvement of this
generalized \Msbar\ result over the zero-mass \Msbar\ result is
encouraging, since the generalized \Msbar\ formalism represents a more
accurate formulation of PQCD. However, a difference of $\chi^{2}$ of 29 is
within the current estimated range of uncertainty of PDF analysis.
\cite{cteq6,Hesse,Lagrange} Therefore, the significance of this difference
is arguable. We also note that the improvement in $\chi^{2}$ is spread over
most of the data sets: there is no smoking gun for the overall difference.

The last two columns of Table~\ref{tbl:fits}\ compare the above results
with two possible uses of the PDFs that represent a \emph{mis-use} of PQCD
in principle, but occur frequently in the literature in practice, perhaps
out of necessity. These involve using PDFs obtained in the general-mass
scheme convoluted with hard-scattering cross-sections (Wilson coefficients)
defined in the zero-mass scheme, and vice versa.%
\footnote{For instance, the MRST distributions \cite{MRST} are obtained in
the General-Mass formalism (using the Robert-Thorne implementation
\cite{ThorneRob}); they are often used in applications convolving with
readily available Zero-Mass hard-scattering cross-sections.} %
For the same data sets, these mis-matched  schemes  result in a difference
of 420$\sim$490 in the overall $\chi^{2}$. These are quite large
differences relative to the tolerances discussed in
Refs.~\cite{cteq6,MRST}, and results in clear discrepancies with some of
the precision DIS data sets, as can be seen in Table~\ref{tbl:fits}. The
lesson is clear: \emph{for quantitative applications, it is imperative to
maintain  consistency between the PDFs and the hard-scattering
cross-sections}.

%%%%%%%%%%%%%%%%%%%%%%%%%%%%%%%%%%%%%%%%%%%%%%%%%%%%%%%%%%%%%%%%%%%
\section{Comparison with data}

In Ref.~\cite{cteq6}, we presented an extensive comparison between the
CTEQ6 results and the data sets used in the global analysis, including new
ways to explicitly account for the correlated experimental systematic
errors. Since the new C6HQ fit is generally similar to the C6M fit, we
shall not duplicate the same comparisons for those quantities where the
differences are minimal. In subsection~\ref{sec:ChmHera}, we compare with
the DIS charm production data, which were not used in the previous CTEQ6
analysis. In subsection~\ref{sec:CompNu}, we discuss some implications of
the neutrino data.

%%%%%%%%%%%%%%%%%%%%%%%%%%%%%%%%%%%%%%%%%%%%%%%%%%
\subsection{DIS Charm Production Data from HERA}
\label{sec:ChmHera}

Because  the H1 and ZEUS  $F_{2}^{c}$ structure function are sensitive to
the charm and  gluon distributions, these will play an important role in
extracting the C6HQ PDFs. In  Fig.~\ref{fig:Fc}, we display comparisons of
the charm production data from  H1 \cite{H1f2c01} and ZEUS \cite{ZeusF2c}
with the C6HQ theory value. For simplicity,  we combine all the data points
from each experiment in a single plot, and we scale the plot by the
theoretical calculation.  On the x-axis, the data points are ordered using
$x$ and $Q$ as the primary and secondary sorting variables, respectively.
The error bars represent the statistical and uncorrelated systematic errors
added in quadrature. We see that the fits are good with $\chi^{2}$ per data
point of 0.881 (7/8) for H1, and 0.98 (18/18) for ZEUS, respectively.

%%%%%%%%%%%%%%%%%%%%%%%%%%%%%%%%%%%%%%%%%%%%%%%%%%%%%%%%%%%%%%%%%%%
\figFc
%%%%%%%%%%%%%%%%%%%%%%%%%%%%%%%%%%%%%%%%%%%%%%%%%%%%%%%%%%%%%%%%%%%

%%%%%%%%%%%%%%%%%%%%%%%%%%%%%%%%%%%%%%%%%%%%%%%%%%%%%%%%%%%%%%%%%%%
\subsection{Neutrino Data in Global Parton Analyses}
\label{sec:CompNu}

The NuTeV measurement of the weak mixing angle $\sin^2 \theta_{\rm W}$
(Ref.~\cite{nuanom}) has recently focused considerable attention on the
neutrino induced DIS process. \cite{nuint} In this measurement,  the weak
mixing angle $\sin^2 \theta_{\rm W}$ is extracted from the ratio of neutral
current (NC) to charged current (CC) $\nu$ and $\bar{\nu}$ cross-sections.
Even before this  $\sin^2 \theta_{\rm W}$ measurement, there have  been
other long-standing unresolved issues in comparing CC structure functions
$F_2^{CC}$ (measured in $\nu$ and $\bar{\nu}$ scattering) with NC structure
functions $F_2^{NC}$ (measured in $e^\pm$ and $\mu^\pm$ scattering)  at
modestly low-$x$ ($\sim 10^{-2}$). One manifestation of this is that the
recently measured structure function $\Delta x F_3^{\nu/\bar{\nu}N}$ is not
compatible with the QCD predictions. \cite{kosty,dimuons} As this structure
function is particularly sensitive to the heavy quark components, $\Delta x
F_3^{\nu/\bar{\nu}N} \simeq s(x)-c(x)$, it is certainly important to
properly treat the heavy quark mass. While the deviations of $\Delta x
F_3^{\nu/\bar{\nu}N}$ are at a non-dramatic $\sim 1 \sigma$ level,  the
pattern has been  systematic and  persistent.

Therefore, we  re-examine the influence of the neutrino structure functions
in the current global analysis keeping in mind the following issues. First,
the high statistics neutrino structure function data of CCFR used here are
from the newer ``physics model independent'' data analysis \cite{pmi},
rather than the earlier ones which contained model-dependent corrections.\
\cite{ccfrsf} \ Secondly, meaningful (i.e., quantitative) comparison of NC
and CC structure functions (which represent different hard-scattering
processes) can be made only within a common underlying theoretical
framework with an accuracy comparable to the experimental
uncertainties.\footnote{For example, direct comparisons of the ratio
$F_2^\nu/F_2^\mu$ to theory (suggested by the ``5/18-rule'') are meaningful
only at the LO parton model level, which is not appropriate for the
precision data from current experiments.} The  proper treatment of charm
mass effects, such as those described in the current study, are an
important part of that theoretical formalism, cf.~\cite{nuc}.

First, we ask whether the neutrino structure function, $F_2^{\nu/\bar{\nu}
N}$, obtained by the newer model-independent analysis is consistent with
the muon-induced  NC fixed-target structure functions, $F_2^{\mu^\pm N}$,
and the HERA collider structure functions,  $F_2^{e^\pm N}$, in a
consistent NLO analysis.%
\footnote{Ref.~\cite{pmi} suggests these data are consistent by examining
the ratio of the structure functions. However, as explained in the text, it
is not appropriate to compare the ratio of two different structure
functions with theory in QCD beyond LO; a complete analysis involving a
NLO fit is required.} %
Fig.~\ref{fig:f2nu} shows the comparison between the data with the CTEQ6HQ
fit  (using $m_c=1.3\ {\rm GeV}$). \figCcfr We see that, even if this data
set is included in the global fit (cf.,~Table~\ref{tbl:fits}), the measured
values in the low-$x$ bins (left figure) consistently lie above the theory
curves. The systematic deviation of data points from the global fit is
reflected in the relatively large $\chi^2$'s associated with this data set,
as seen in Table~\ref{tbl:fits}. The fit is dominated by the much more
extensive NC data (from the BCDMS, NMC, H1, and ZEUS experiments), hence
one can regard the theory curves as representing the NC data properly
``corrected'' for NLO QCD effects. The pattern of deviation of the neutrino
data from the average NC predictions at low-$x$ is similar to earlier
comparisons. In other words, the discrepancy between CCFR $F_2^{\nu}$ and
the other NC $F_2^{e,\mu}$ measurements at small $x$ persists in the
context of contemporary global QCD analyses, even when charm mass effects
are properly treated as in the present analysis. Other possible theoretical
sources for this difference (in addition to the NLO contributions included
in this comparison) have been discussed in Refs.~\cite{kosty,nuint,nuref},
and in the literature quoted therein. On the experimental side, preliminary
NuTeV measurements \cite{bernstein} seem to yield improved agreements  with
the NLO QCD predictions at low-$x$. It remains to be seen whether the final
NuTeV results will resolve this problem.

In our analysis, the strangeness distribution is constrained to be
proportional to $[\bar{u}(x)+\bar{d}(x)]$; and its normalization at the
scale $Q_0$ is constrained to be
\begin{equation}
\kappa =
{{\int d x x [s(x) + {\bar s}(x) ] } \over
{\int dx x [\bar{u}(x) + \bar{d}(x)] }
}
\  \simeq \ {1 \over 2}
\quad ,
\label{eq:kappa}
\end{equation}
as in most current global analysis.  In the future, a NLO analysis of the
CC charm production data ($\nu s \to c \mu \to \mu^\pm \mu^\mp X$) from
CCFR and NuTeV \cite{charmcc} has the potential to determine $\{ s(x),
{\bar s} (x)\}$ with much more precision.  Then, a re-assessment of the
apparent discrepancy between the measured  $\Delta x F_3^{\nu/\bar{\nu}N}$
and the low-lying QCD predictions \cite{kosty} will be warranted. An
accurate determination of the strangeness of the proton, as well as the
strangeness asymmetry  $[s(x)-{\bar s}(x)]$ will have important
implications for these measurements, including the NuTeV anomaly, cf.,
Refs.~\cite{nuanom,nuint}.

Clearly, the interpretation of the neutrino DIS data requires a detailed
NLO analysis. Global QCD analyses of parton distributions, such as the one
performed in this paper, are essential to  unraveling the underlying
physics.  Progress on both the experimental and the theoretical fronts are
needed;  this advancement is ongoing.

%%%%%%%%%%%%%%%%%%%%%%%%%%%%%%%%%%%%%%%%%%%%%%%%%%%%%%%%%%%%%%%%%%%
\section{Comparison with related PDFS: CTEQ6HQ, CTEQ6M, CTEQ5HQ}

The CTEQ6HQ and CTEQ6M fits provide comparable descriptions of the global
QCD data in two \textit{different} schemes. Some of the  differences in the
PDFs arise purely from the choice of scheme. We are particularly interested
in the differences for the charm distribution, and the closely correlated
gluon distribution, due to the improved treatment of heavy quark effects in
the generalized \Msbar\ scheme. It is also interesting to compare the
differences between the earlier (previous generation) CTEQ5HQ (C5HQ)
distributions with the new CTEQ6HQ distributions; differences between these
PDFs are attributable both to new data, and to minor differences in the way
the theoretical inputs are implemented. We combine these comparisons in the
figures presented below.

%%%%%%%%%%%%%%%%%%%%%%%%%%%%%%%%%%%%%%%%%%%%%%%%%%%%%%%%%%%%%%%%%%%
\subsection{Light partons at the input scale}

Figs.~\ref{fig:UDqkA}a,b show the $u$-quark and $d$-quark distributions
from C5HQ, C6M and C6HQ at the initial $Q$-scale of 1.3~GeV (the charm
mass). In order to exhibit the behavior of the PDFs at both large and small
$x$ values, the x-axis is scaled linearly in $\sqrt{x}$. To give some
additional physical insight to these plots, the y-axis is chosen to be
$x^{1.5}f(x,Q)$, so that the area under each curve represents the momentum
fraction carried by that distribution. This chosen scaling then focuses on
the valence peak, and suppresses the small-$x$ sea excitations. We see that
the valence $u$-quark and $d$-quark distributions are well constrained by
the precision DIS data;  the scheme dependence is not pronounced.

\figUDqkA
\figUDqkB

Figs.~\ref{fig:UDqkB}a,b show the same PDFs, except the axes are scaled
differently in order to highlight the sea quark distributions. The
differences between the CTEQ5 and CTEQ6 generations of PDFs are a result of
the improved low-$x$ data. When we compare the HQ and \Msbar\ type PDFs of
the CTEQ6 generation, we observe the general pattern that the HQ
distributions overshoot the \Msbar\ ones at low $x$ ($\sim 10^{-3}$), and
that the distributions meet again at ultra-small $x$  ($\lsim  10^{-5}$).
We can qualitatively understand the  larger sea quark distributions for the
HQ scheme since these distributions are compensating for the suppression of
charm in this scheme; this suppression is absent in the zero-mass \Msbar\
scheme. In the region of  ultra-small $x$, there is no data constraining
these distributions;  we show this effect graphically in the next Section
where we consider the PDF  uncertainty band of the the HQ and \Msbar\
distributions.

\figGluA

Fig.~\ref{fig:GluA} shows the comparison of the gluon distribution at
$Q_{0}$. Here the difference between C5HQ and the CTEQ6 generation of gluon
distributions is pronounced. As discussed in Ref.~\cite{cteq6}, the change
in this least-well-determined parton distribution is due to the recent
precision DIS data (most influential in the small $x$ region) in
conjunction with the greatly improved inclusive jet data from the Tevatron
(critical for the medium to large $x$ regions). The differences between
C6HQ and C6M gluons at large $x$ are due to a combination of
scheme-dependence, and the inherent uncertainty range of the current
analysis.\ \cite{cteq6}

Fig.~\ref{fig:StrA} shows the comparison of the strange distributions at
the same $Q_{0}$.  The noticeable difference between the C5HQ curve and the
others, in this case, is largely the result of different theoretical
inputs: namely, the  $\kappa$ parameter  (cf., Eq.~\ref{eq:kappa}). $\kappa
$ determines  the ratio of strange to non-strange sea quarks  at the
initial scale $Q_{0}$.  This $\kappa $ factor, known only approximately,
was chosen to be $1/2$ both in the CTEQ5 and CTEQ6 analyses, but for
slightly different values of $Q_{0}$---1.0 GeV for C5HQ and 1.3 GeV for
CTEQ6 sets.

\figStrA

%%%%%%%%%%%%%%%%%%%%%%%%%%%%%%%%%%%%%%%%%%%%%%%%%%%%%%%%%%%%%%%%%%%
\subsection{Charm and gluon distributions at $Q^2=10$ GeV$^{\mathrm{2}}$}

\figChmGlu

To compare the differences of the charm distributions, we need to move some
distance above the charm mass scale, since all current sets assume a zero
charm distribution at $\mu=m_c$. Fig.~\ref{fig:ChmGlu}a makes this
comparison at $Q^{2}=10$ GeV$^{\mathrm{2}}$. We see a substantial
difference between the C5HQ and the two CTEQ6 charm distributions. Unlike
the case of the strange quark, this difference is physical; it is mainly a
reflection of the difference in the gluon distributions (shown already for
a lower $Q$ value in Fig.~\ref{fig:GluA}) because charm is radiatively
generated from the gluon. To confirm this hypothesis,
Fig.~\ref{fig:ChmGlu}b shows the gluon distributions at the same $Q^{2}=10$
GeV$^{\mathrm{2}}$; the similarity is clear.

%%%%%%%%%%%%%%%%%%%%%%%%%%%%%%%%%%%%%%%%%%%%%%%%%%%%%%%%%%%%%%%%%%%
\subsection{$Q$-dependence of heavy quark mass effects}

In this subsection, we compare the $Q$ dependence of the C6M and C6HQ
parton distributions at fixed values of $x$ in order to see how the
differences between PDFs in the zero-mass and general-mass schemes vary
with increasing $\mu$-scale. We expect these differences (relative to the
general PDF uncertainties) will decrease with increasing $Q$. The purpose
of this study is two-fold. First, we check the self-consistency of the
analysis, which is based on the expectation that power suppressed mass
terms vanish asymptotically as $Q \rightarrow \infty$. Secondly, we
determine the specific scales where the  mass-effects become insignificant,
in practice. The latter question is relevant for high-energy applications
of PDFs derived from lower-energy data.%
\footnote{It also serves to validate the self-consistency of analyses where
threshold effects have been neglected  (such as for collider processes).} %
 It can also be seen as a general illustrative example demonstrating how
the relative importance of power suppressed terms decrease in the high $Q^2$
region where the conventional zero-mass PQCD
results becomes dominant.

To examine the effects due to heavy flavor masses in the theoretical
formalism, we cannot directly compare  physical quantities that are being
fit---they are the same (within the errors of the global fit) by
construction. On the other hand, parton distribution function of individual
flavors are not a good gauge either, because they are not physical---they
are scheme dependent.  Thus, for this comparison, we compromise and examine
the quark singlet combination ($\Sigma \equiv \sum_{q} (q+{\bar q})$) 
and the gluon distribution.

 \figQdep

Figs.~\ref{fig:Qdep}a,b show, as solid lines, the C6HQ gluon and quark
singlet distributions normalized to that of C6M. The shaded bands represent
the ranges of uncertainty of the relevant quantities due to experimental
sources, as estimated in the CTEQ6 global analysis.\ \cite{cteq6} We
observe the following pattern: for most of the parameter space, the
deviation of the C6HQ from C6M falls within the estimated uncertainty
range. In particular, for the gluon distribution this is true for all
values of $x$.
 For the quark distribution, the HQ fit overshoots the zero-mass fit
significantly (i.e., beyond the uncertainty band) at moderately low
$x$ $\sim [10^{-2} , 10^{-3}]$.
At very low $x$ $\sim 10^{-5}$, the differences fall
within the uncertainty bands again (primarily due to the increasing
uncertainty). Although the HQ quark PDFs at $x \sim [10^{-2} , 10^{-3}]$
eventually evolve into the error band of the zero-mass fit as $Q$
increases, one may wonder why the two fits do not merge faster.
 The GM-VFNS and ZM-VFNS differ by the manner in which they organize
terms of order ${\cal{O}}(m^2/Q^2)$. Thus, the Wilson coefficients
in the two schemes
will be comparable when $(m^2/Q^2) << 1$.
The boundary conditions of the PDFs at $Q_0$ will also differ between the
two schemes by terms of order  ${\cal{O}}(m^2/Q^2)$; however, since the
 PDFs  evolve
logarithmically in the scaling variable, while the power corrections
fall off more quickly, this can contribute to scheme differences
at scales significantly above the scale $m$. This effect is evident
in the case of the singlet quark PDF at moderately low $x$ values.
 Thus, the observed behavior is quite natural.

%%%%%%%%%%%%%%%%%%%%%%%%%%%%%%%%%%%%%%%%%%%%%%%%%%%%%%%%%%%%%%%%%%%
\section{Concluding Remarks}

The CTEQ6HQ PDFs presented here complement the previously published CTEQ6
sets by providing distributions which can be used in the generalized MS-bar
scheme with non-zero mass partons.
 This analysis includes the complete set of NLO processes including
the real and virtual quark-initiated terms. Additionally, the ACOT($\chi$)
scheme is used to introduce a generalized scaling variable which provides
numerically stable results for the entire energy range---from heavy quark
thresholds to the high energy limit.

While the zero-mass parton scheme is sufficient for many purposes, the
fully massive scheme can be important when physical quantities are
sufficiently sensitive to heavy quark contributions. This is evident when
comparing the CTEQ6HQ and CTEQ6M fits to the mis-matched sets (Table~1)
where the precise DIS data from HERA highlights the discrepancies.

The CTEQ6HQ fits also provide the basis for a series of further studies
involving more quantitative analysis of strange, charm, and bottom
quark distributions inside the nucleon.
For example, the CTEQ6HQ PDFs are necessary for a consistent analysis
of resummed differential distributions for heavy quark production
such as in Ref.~\cite{Nadolsky:2002jr}.
 Using the full range of data from both the charged and neutral current
processes, these distributions can reduce the uncertainties in the
calculations; hence, they  have significant implications for charm and
bottom production, and can help resolve questions about intrinsic heavy quark
constituents inside the proton, the $\Delta x F_3$ structure
function, and the extraction of $\sin\theta_W$.

%%%%%%%%%%%%%%%%%%%%%%%%%%%%%%%%%%%%%%%%%%%%%%%%%%%%%%%%%%%%%%%%%%%
\section*{Acknowledgment}

We thank our colleagues J.~Huston, P.~Nadolsky, J.~Pumplin, and D.~Stump,
for fruitful collaboration on the CTEQ6 project that forms the foundation
of this study.  F.O.\ acknowledges the hospitality of MSU and BNL where a
portion of this work was performed.
S.K.~is grateful to RIKEN, Brookhaven National Laboratory and
the U.S.~Department of Energy (contract No.~DE-AC02-98CH10886)
for providing the facilities essential for the completion of this work.
This research was supported by the National
Science Foundation (grant No.~0100677), and by the Lightner-Sams
Foundation.

%%%%%%%%%%%%%%%%%%%%%%%%%%%%%%%%%%%%%%%%%%%%%%%%%%%%%%%%%%%%%%%%%%%
% References

%%%%%%%%%%%%%%%%%%%%%%%%%%%%%%%%%%%%%%%%%%%%%%%%%%%%%%%%%%%%%%%%%%%
\end{document}